\documentclass[conference]{IEEEtran}
\IEEEoverridecommandlockouts
\usepackage{cite}
\usepackage{amsmath,amssymb,amsfonts}
\usepackage{algorithmic}
\usepackage{graphicx}
\usepackage{textcomp}
\usepackage{xcolor}
\usepackage{comment}
\def\BibTeX{{\rm B\kern-.05em{\sc i\kern-.025em b}\kern-.08em
    T\kern-.1667em\lower.7ex\hbox{E}\kern-.125emX}}
\usepackage{fancyhdr}
\usepackage{subcaption}

\begin{document}

\title{Ternary and Quaternary CNTFET Full Adders are less efficient than the Binary ones for Carry-Propagate Adders\\
}

\author{\IEEEauthorblockN{ Daniel Etiemble}
\IEEEauthorblockA{\textit{LISN} \\
\textit{University Paris Saclay}\\
Gif sur Yvette, France \\
de@lri.fr}

}

\maketitle

\begin{abstract}
In Carry Propagate Adders, carry propagation is the critical delay. The most efficient scheme is to generate $C_{out0}$ ($C_{in}$=0) and $C_{out1}$($C_{in}$=1) and multiplex the correct output according to $C_{in}$. For any radix, the carry output is always 0/1.
We present two versions of ternary adders with $C_{in}$=(0V, $V_{dd}$/2) and $C_{in}$ = (0V, $V_{dd}$) and two versions of quaternary adders with $C_{in}$=(0V, $V_{dd}$/3) and $C_{in}$ = (0V, $V_{dd}$). Using full swing $V_{dd}$ for $C_{in}$ reduces the propagation delays for ternary and quaternary adders. 6-bit, 4-trit and 3-quit CPAs are then compared.
\end{abstract}


\section{Introduction}
\begin{figure}[htbp]
\centerline{\includegraphics[width=8cm]{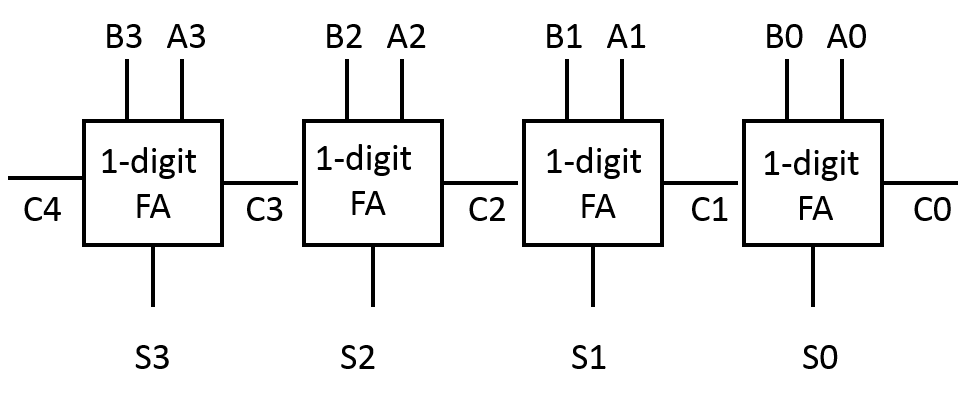}}
\caption{4-digit Carry Propagate Adder}
\label{CPA4}
\end{figure}

Carry Propagate Adders (CPAs) are the most simple N-digit adders. Figure \ref{CPA4} presents a 4-digit CPA. Whatever digit radix is used (2,3,4...), the carries are always 0/1. In this paper, we consider both binary, ternary and quaternary CPAs. From Figure \ref{CPA4}, it results that the performance of a CPA is a direct function of the used 1-digit full adder. More precisely, the critical delay path of a CPA is related to the carry propagation.

\subsection{Carry values in a m-valued adder}

The truth table of the ternary full adder is shown in Table \ref{3TT}. The truth table of the quaternary full adder is shown in Table \ref{T1}. In both cases, the input and output carries are binary (0,1). This property is valid for any radix. So, ternary 1-trit full adders have ternary inputs and output (0,1,2) and binary carries (0,1). Quaternary 1-digit adders have quaternary inputs and outputs (0,1,2,3) and binary carries (0,1).
There are two techniques to get the different levels:
\begin{itemize}
\item Using two power supplies $V_{dd}$ and $V_{dd}$/2 for ternary circuits and three power supplies $V_{dd}$, $2V_{dd}$/3 and $V_{dd}$/3 for quaternary circuits. 
\item Using only one power supply ($V_{dd}$) and get the intermediate values through transistor connected as resistors. In that case, there is a large static power dissipation resulting from the direct current flow through the voltage divider for intermediate levels. This is why we only consider the option with two or three power supplies.
\end{itemize} 
With this approach, all proposed designs use $V_{dd}$/2 as the voltage when carry=1 for ternary circuits and $V_{dd}$/3 for quaternary ones . This  raises one question : must the binary carries mandatory use 0 and 1 ternary/quaternary values or can they use 0/2 ternary values or 0/3 quaternary values? In other words, 0 and $V_{dd}$/2 or 0 and $V_{dd}$ in the ternary case and 0 and $V_{dd}$/3 or 0 and $V_{dd}$  when the ternary/quaternary adders have a $V_{dd}$ power supply. In this paper, we consider ternary and quaternary circuits using the two different approaches for carry levels and we compare them with binary adders. The comparison is extended to CPAs computing approximately the same amount of information:  6-bit (Binary Digit) CPAs, 4-trit (Ternary Digit) CPAs and  3-quit (Quaternary Digit) CPAs.
The paper is organized as follow:
\begin{itemize}
\item we present the methodology
\item we present the different ternary adders and their performance
\item we present the different quaternary adders and their performance
\item We present the different binary adders that are used for comparison with their performance
\item we compare the performance of the quaternary, ternary and binary CPAs computing the same amount of information
\item we finally conclude.
\end{itemize}

\begin{table}
\centering
\caption{Truth table of a ternary full adder}
\begin{tabular}{|cc|cc||cc|cc|}
\hline
 \multicolumn{4}{|c||}{$C_{in}$=0} & \multicolumn{4}{|c|}{$C_{in}$=1}\\
  \hline
 A&B&$S_0$&$C_{out0}$&A&B&$S_1$&$C_{out1}$\\
\hline 
 0&0&0&0&0&0&1&0\\
 0&1&1&0&0&1&2&0\\
 0&2&2&0& 0&2&0&1\\
 1&0&1&0&1&0&2&0\\
 1&1&2&0&1&1&0&1\\
 1&2&0&1&1&2&1&1\\
 2&0&2&0&2&0&0&1\\
 2&1&0&1&2&1&1&1\\
 2&2&1&1&2&2&2&1\\
  \hline
\end{tabular}
\label{3TT}
\end{table}

\begin{table}
\centering
\caption{Truth table of a quaternary adder}
\begin{tabular}{|c|c|c||c|c|c|c|c|c||c|c|}
  \hline
A&B&$C_{in}$&$S_0$&$C_{out0}$& &A&B&$C_{in}$&$S_0$&$C_{out1}$\\
\hline
 0&0&0&0&0&&0&0&1&1&0\\
0&1&0&1&0&&0&1&1&2&0\\
0&2&0&2&0&&0&2&1&3&0\\
0&3&0&3&0&&0&3&1&0&1\\

 1&0&0&1&0&& 1&0&1&2&0\\
1&1&0&2&0&&1&1&1&3&0\\
1&2&0&3&0&&1&2&1&0&1\\
1&3&0&0&1&&1&3&1&1&1\\

 2&0&0&2&0&& 2&0&1&3&0\\
2&1&0&3&0&&2&1&1&0&1\\
2&2&0&0&1&&2&2&1&1&1\\
2&3&0&1&1&&2&3&1&2&1\\

 3&0&0&3&0&&3&0&1&0&1\\
3&1&0&0&1&&3&1&1&1&1\\
3&2&0&1&1&&3&2&1&2&1\\
3&3&0&2&1&&3&3&1&3&1\\
  \hline
\end{tabular}
\label {T1}
\end{table}

\subsection{Related works}
A lot of ternary full adders have been published in the last decade \cite{R1,R2,R3,R4,R5,R6,R7,R8,R9}. They use different techniques. Transistor count is not a sufficient criteria to determine the best TFAs.  However, considering Table \ref{tfa} and a similar Table comparing Ternary Half Adders in \cite{R10}, we may consider that the technique using unary operators and MUXes is the most efficient one.
Several quaternary full adders with CNTFET simulations have been published in the last decade \cite{R12,R13,R14}. In a preprint paper without simulations, several possible implementations based on the transistor count have been evaluated\cite{R15}. This paper considers CPAs.

\begin{table}[t!]
\caption{TFAs Comparison}
\label{tfa}
\setlength{\tabcolsep}{4pt}
\centering
\begin{tabular}{c|c|l}
\hline
					&CNTFETs	& Technique \\
	TFA / Year		&Count	&  \\\hline
In \cite{R1} 2011	&412& Decoders-Binary-Encoder\\
In \cite{R2} 2017	&105		& Two custom algorithm + TMuxes\\
In \cite{R3} 2017	&74		& TMUXes\\
In \cite{R4} 2018	&89		& TMUXes\\
In \cite{R5} 2018	&98		& TBDD algorithm\\
In \cite{R6} 2019	&142	&Unary ops +MUXes+Encoder	\\
In \cite{R7} 2020	&74 & Pass transistors + MUXes\\
In \cite{R8} 2020	&106	&Modified Quine-McCluskey algorithm\\
In \cite{R9} 2021	&54 & Unary ops + Decoders + Transmission gates\\
\hline
\end{tabular}
\end{table}

\subsection{Contributions}
This paper compares efficient implementations of CNTFET binary, ternary and quaternary adders.
The main contributions of this paper are
\begin{itemize}
\item For ternary and quaternary adders, we use both the carry swing corresponding to 0 and 1 usual carry swing and the full $V_{dd}$ carry swing. Using full carry swing reduces the carry propagation delays.
\item While ternary and quaternary adders use a $V_{dd}$ power supply, we consider binary adders with both $V_{dd}$ and $V_{dd}$/2 power supplies. Reduced voltage swings for binary adders drastically reduce power dissipation and the Power Delay Product (PdP).
\item We compare the performance of 6-bit CPAs, 4-trit CPAs and  3-quit CPA that computes the same or approximately the same amount of information.
\end{itemize}

\section{Methodology}  
The significant figures to compare circuit designs include switching times, power dissipation, chip area, etc. The comparison is realized by using HSpice simulations and evaluating the chip area according to transistor sizes.
\subsection{CNTFET technology}
All simulations are done with the 32nm CNTFET parameters of Stanford library \cite{R16}. 
We use CNTFET technology for two main reasons:
\begin{itemize}
\item Simulations parameters for the most recent FinFET technologies are not available
\item Most of papers presenting designs of ternary or quaternary circuits in the last period use simulations with this 32 nm CNTFET technology. This allows our results to be compared with all published results on ternary or quaternary circuits.
\end{itemize}
One advantage of CNTFET technology is that the threshold levels of gates only depend on the diameter of individual transistors, which facilitates the design of m-valued circuits. 

\subsection{Propagation delays}
Generally, propagation delays are presented as an average of the delays corresponding to all combinations of input transitions. This presentation could be confusing. For the CPA presented in Figure \ref{CPA4}, $A_i$, $B_i$ and $C_0$ inputs are simultaneously available. The important information is the propagation delay corresponding to the critical paths, i.e. from $C_0$ (or $A_0$/$B_0$) to $C_4$ and $S_3$. When the 4-digit CPA is used to build larger CPAs, the critical path is from $C_0$ to $C_4$. 
We will only present the propagation delays corresponding to the critical paths.
\subsection{Power and Energy dissipation}
\begin{figure}
  \begin{subfigure}{\linewidth}
  \includegraphics[width=4 cm]{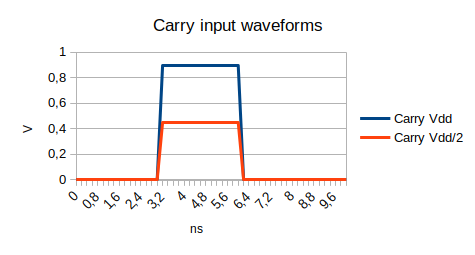}
  \includegraphics[width=4 cm]{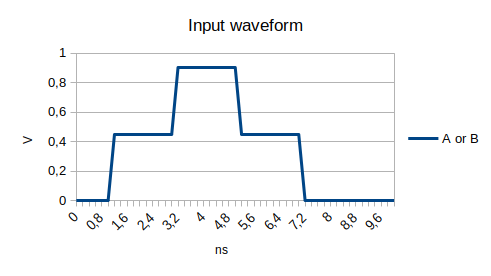}
    \caption{Ternary waveforms}
  \end{subfigure}
   \begin{subfigure}{\linewidth}
  \includegraphics[width=4 cm]{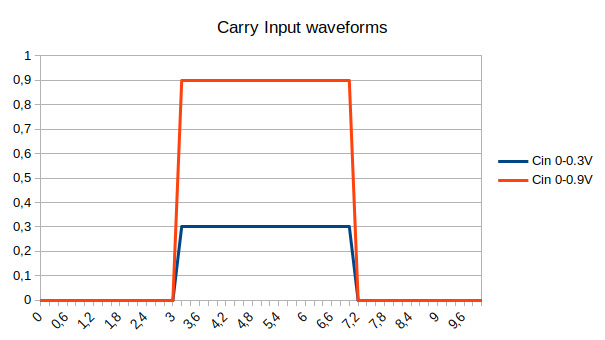}
  \includegraphics[width=4 cm]{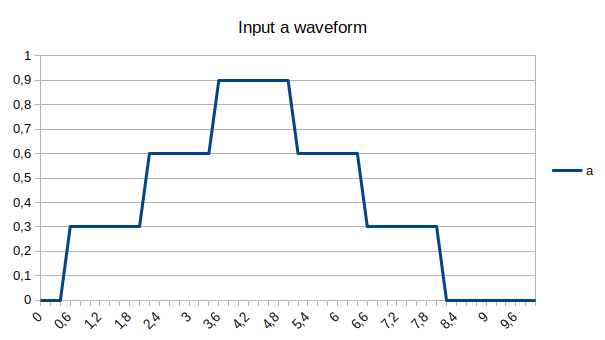}
    \caption{Quaternary waveforms}
  \end{subfigure}
  \begin{subfigure}{\linewidth}
  \includegraphics[width=4 cm]{CTW.png}
  \includegraphics[width=4 cm]{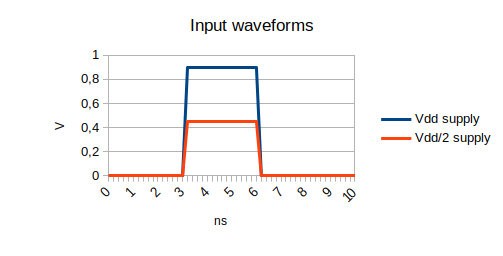}
    \caption{Binary waveforms}
  \end{subfigure}
  \caption{Input waveforms for all simulations}
  \label{WF}
\end{figure}

 Power and PDP (Power Delay Product) directly depends on the duration of the input signals. It is important to use the same input signal for all designs.
For all simulations, we use the input waveforms shown in Fig. \ref{WF}. We have verified that the delays for 0-2 or 2-0 ternary transitions are always less than ternary transitions 0-1, 1-2, 2-1 or 1-0. The situation is similar for quaternary transitions.
We use these waveforms to compute the worst-case delays from Input (A or B) to Sum/$C_{out}$ and from $C_{in}$ to Sum/$C_{out}$.

\subsection{Chip area}
Without drawing the layout of the circuits, there is no technique to evaluate the chip area. 
We use a rough evaluation of the chip area by summing the diameters of all the used transistors by each circuit. This rough evaluation is a little bit better than the transistor count. In this paper, we use the diameter values presented in Table \ref{Diameter}.
\subsection{Circuit styles}
Many techniques have been proposed to design full adders. We only consider techniques with the following properties:
\begin{itemize}
\item No static power dissipation
\item The circuit outputs have full swing. Reduced swings degrade noise margins and can degrade  the operation of cascaded circuits, such as CPAs
\item The circuits should have a sufficient driving capability. This point is outlined in subsection \ref{CP}.
\end{itemize}

\begin{table}
\centering
\caption{Transistor diameters}
\begin{tabular}{|c|c|c|c|c|c|c|c|c|}
  \hline
n&Diameter (nm)&$|Vth|$ (V)\\
  \hline
  8&0.626&0.696\\
10&0.783&0.557\\
13&1.018&0.428\\
19&1.487&0.293\\
29&2.27&0.192\\
37&2.896&0.150\\
  \hline
\end{tabular}
\label {Diameter}
\end{table}

\subsection{Temperature}
All the simulations are done with a 25°C temperature. The ternary adders (section \ref{QA}), the quaternary one (section \ref{QA}) and one of the three binary ones (section \ref{BA}) use the same circuit style. The same CNTFET parameters are used. There are very few opportunities that different temperatures would change the results of the comparisons between the different adders.

\subsection{Carry propagation in Carry Propagate Adders (CPAs)} \label{CP}
As previously mentioned, carry propagation delay is the critical one in CPAs. This delay must be minimized, either   for  quaternary FAs or for ternary FAs or for binary FAs. One technique is illustrated in Figure 3: 
$C_{out}$ = $C_{out0}$ when $C_{in}$=0 and $C_{out}$ = $C_{out1}$ when $C_{in}$=1. 
 The correct $C_{out}$ is obtained via a multiplexer implemented with transmission gates. This technique  is used in many published binary full adders, such as Transmission Gate Adders (TGAs) and other ones quoted in\cite{Wairya}. This approach has a major drawback for CPAs. When there is a direct propagation from the first to the last full adders, there is a RC line effect (Figure \ref{RCeffect}) associated to the capacitive loads that significantly degrade the carry delays. To get a minimal delay without degrading the switching times, the $C_{out}$ signal must be
 restored by an inverter gate, as shown in Figure \ref{Coutbmux}. It means that $\overline{C_{out0}}$ and $\overline{C_{out1}}$ should be computed, transmitted to $\overline{C_{out}}$ through the multiplexer and the final inverter generates $C_{out}$. This technique will be used in most circuits that we present.

\begin{figure}[htbp]
\centerline{\includegraphics[width=3cm]{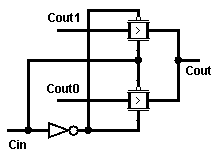}}
\caption{$C_{in}$ to $C_{out}$ carry propagation in a full adder}
\label{Coutmux}
\end{figure}

\begin{figure}[htbp]
\centerline{\includegraphics[width=4cm]{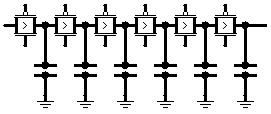}}
\caption{RC effect with series of transmission gates}
\label{RCeffect}
\end{figure}

\begin{figure}[htbp]
\centerline{\includegraphics[width=4cm]{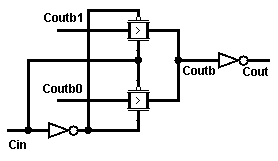}}
\caption{$C_{in}$ to $C_{out}$ carry improved propagation with capacitive loads in a full adder}
\label{Coutbmux}
\end{figure}

\section{The Ternary Full Adders}
\label{TA}
\begin{figure}[htbp]
\centerline{\includegraphics[width=7
cm]{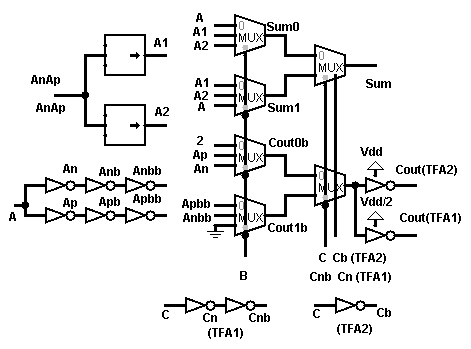}}
\caption{1-trit Full Adder (Mux approach)}
\label{TFA}
\end{figure}

We consider two different ternary full adders based on the MUX approach which common scheme is presented in Fig. \ref{TFA}. The difference between the first one (called TFA1) and the second one (TFA2) are detailed after the presentation of the MUX approach. For TFA1 and TFA2, two versions are considered that differ by the carry voltage swing (0-$V_{dd}$/2) and (0-$V_{dd}$).

\subsection{The MUX approach} 
According to Table \ref{3TT},
When $C_{in}$=0
\begin{itemize}
\item When B=0 then Sum=A
\item When B=1 then Sum = (A+1) mod(3) quoted as $A^{1}$
\item When B=2 then Sum = (A+2) mod(3) quoted as $A^{2}$
\item When B=0 then Carry=0
\item When B=1 then Carry=1 when $A=2$ else 0
\item When B=2 then Carry=1 when $A>0$ else 0
\end{itemize}
When $C_{in}$=1
\begin{itemize}
\item When B=0 then Sum=$A^{1}$
\item When B=1 then Sum=$A^{2}$
\item When B=2 then Sum= A
\item When B=0 then Carry=1 when $A=2$ else 0
\item When B=1 then Carry=1 when $A>0$ else 0
\item When B=2 then Carry=1
\end{itemize}

Functions $A^{1}$ and $A^{2}$ are presented in Table \ref{A12}

\begin{table}
\centering
\caption{Functions $A^{1}$ and $A^{2}$}
\begin{tabular}{|c|c|c|c|c|c|c|c|c|}
  \hline
 &$A^{1}$&$A^{2}$\\
  \hline
0&1&2\\
1&2&0\\
2&0&1\\
  \hline
\end{tabular}
\label {A12}
\end{table}

\begin{table}
\centering
\caption{NI and PI binary functions}
\begin{tabular}{|c|c|c|c|c|c|c|c|c|}
  \hline
 &NI&PI\\
  \hline
0&2&2\\
1&0&2\\
2&0&0\\
  \hline
\end{tabular}
\label{T2}
\end{table}

\subsection{The ternary full adders}

\begin{figure}[htbp]
\centerline{\includegraphics  [width =5 cm]{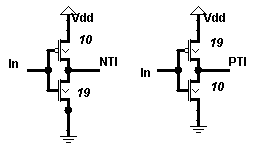}}
\caption{Threshold detectors}
\label{TDEC3}
\end{figure}

\begin{figure}[htbp]
\centerline{\includegraphics[width=8cm]{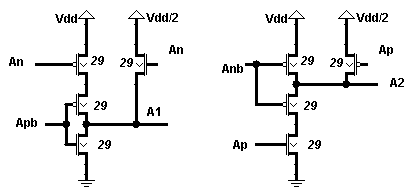}}
\caption{A¹ and A² circuits}
\label{A1A2}
\end{figure}

TFA1 and TFA2 use the same threshold detectors (Fig. \ref{TDEC3}). They implement the NI (Negative Inverter) and PI (Positive Inverter) functions presented in Table \ref{T2}.  
The operators $A^{1}$ and $A^{2}$  are derived from the threshold detectors as shown in Fig. \ref{A1A2}. $A_n$ is the output of a negative inverter, $A_p$ is the output of a positive inverter. $A_{nb}$ and $A_{pb}$ are the outputs of binary inverters with inputs $A_n$ and $A_p$.

\begin{figure*}[htbp]
\centerline{\includegraphics[width=12cm]{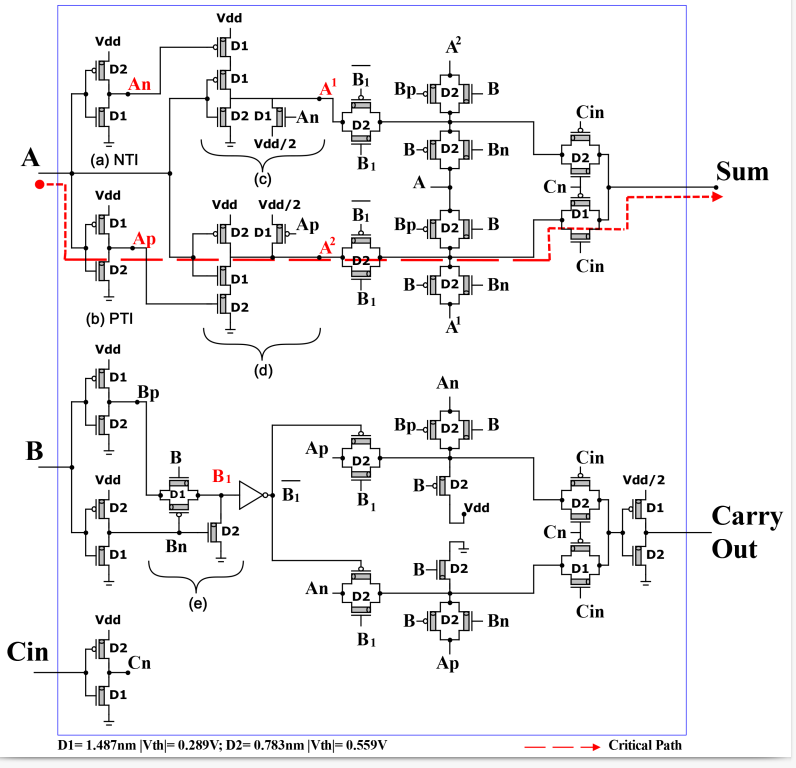}}
\caption{Specific implementation of ternary MUXes for Ternary Full Adder \cite{Jaber20213}}
\label{TFA55S}
\end{figure*}

\begin{figure}[htbp]
\centerline{\includegraphics[width=6cm]{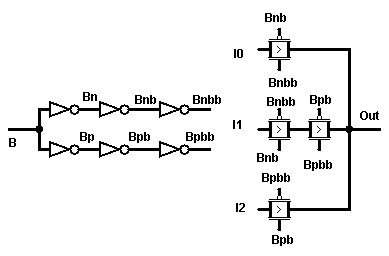}}
\caption{3-input MUX with ternary control}
\label{TMUX3}
\end{figure}

TFA1 and TFA2 differ by the implementation of the MUX operators. TFA1 has a specific implementation of sum MUX and carry MUX \cite{Jaber20213} in the ternary adder (Fig. \ref{TFA55S}). TFA2 uses the MUX3 operators shown in Fig. \ref{TMUX3}. TFA1 and TFA2 use the same MUX2  implementation. 

There are few differences between 0.45V and 0.9V carry versions. TFA1 uses a NTI  inverter to get $C_{in}$ and the final carry inverter has a 0.45V power supply. For TFA2, $C_{in}$ and $C_{out}$ use 0.9V inverters. TFA1 $\Sigma(Di)$ = 72 nm (for carry swing = $V_{dd}$/2 = 0.45V) and $\Sigma(Di)$ = 73 nm (for carry swing = $V_{dd}$= 0.9V). TFA2 $\Sigma(Di)$ = 111 nm for carry swing = 0.45V and $\Sigma(Di)$ = 112 nm for carry swing = 0.9V.

\subsection{Performance with a 2 fF capacitive load}
\begin{figure*}[htbp]
\centerline{\includegraphics[width=12 cm]{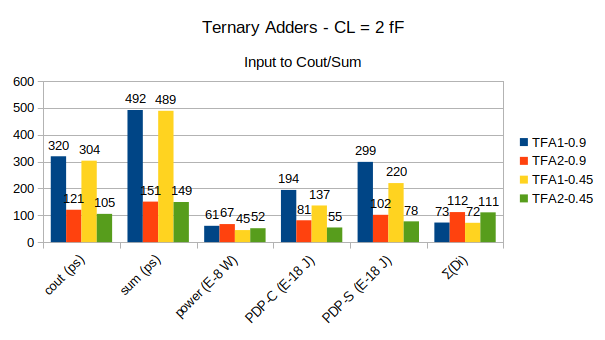}}
\caption{Input to $C_{out}$/Sum performance of ternary adders}
\label{CTI20}
\end{figure*}

\begin{figure*}[htbp]
\centerline{\includegraphics[width=12 cm]{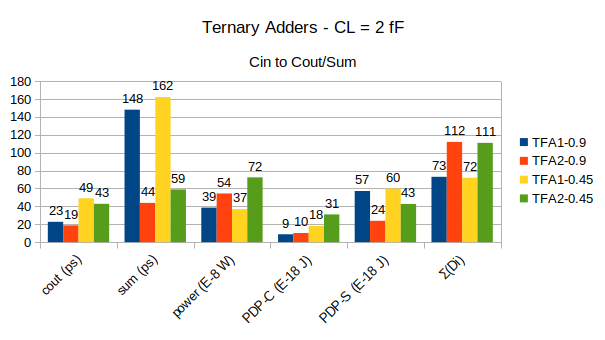}}
\caption{$C_{in}$ to $C_{out}$/Sum performance of ternary adders}
\label{CTC20}
\end{figure*}

Fig. \ref{CTI20} presents the Input to $C_{out}$/Sum performance with a $C_{L}$ = 2 fF capacitive load. Fig. \ref{CTC20} presents the $C_{in}$ to $C_{out}$/Sum performance with the same load.

The following remarks can be made
\begin{itemize}
\item TFA2 has a larger $\Sigma(Di)$ (x1.5).
\item There is a huge difference for Input to $C_{out}$/Sum. The only difference comes from the implementation of ternary MUXes. MUX3 implementation of TFA2 is more efficient.
\item There is little difference in $C_{in}$ to $C_{out}$ delay for TFA1 and TFA2. This is not surprising as both uses similar MUX2 + Inverter designs for this propagation. 
\item TFA1 and TFA2 with 0.9V carry swing have $C_{in}$ to $C_{out}$ delay roughly two times faster than TFA1 and TFA2 with 0.45V carry swing. The 0.9V inverters have more driving capability than the 0.45V inverters. 
\end{itemize} 
While TFA2 has 50\% more $\Sigma(Di)$, the huge difference in Input to Sum delay for the last stage of a CPA makes TFA2 the best ternary adder either with 0.45V or 0.9V carry swing.

\subsection{Delays and power according to capacitive load}

\label{TR}
\begin{figure}[htbp]
\centerline{\includegraphics[width=9cm]{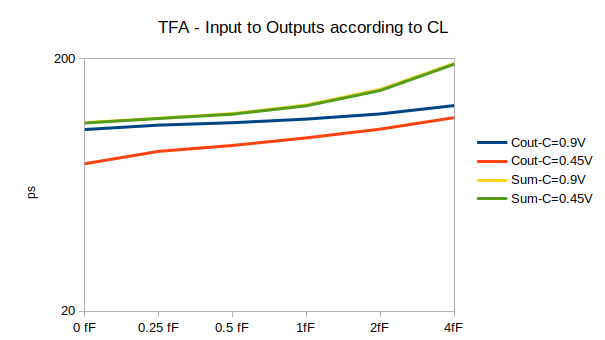}}
\caption{TFA-Input to $C_{out}$/Sum delays according to $C_{L}$}
\label{TI20CL}
\end{figure}

\begin{figure}[htbp]
\centerline{\includegraphics[width=9cm]{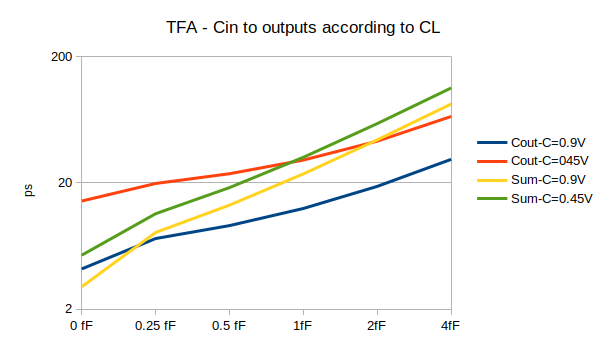}}
\caption{TFA-$C_{in}$ to $C_{out}$/Sum delays according to $C_{L}$}
\label{TC20CL}
\end{figure}

\begin{figure}[htbp]
\centerline{\includegraphics[width=9cm]{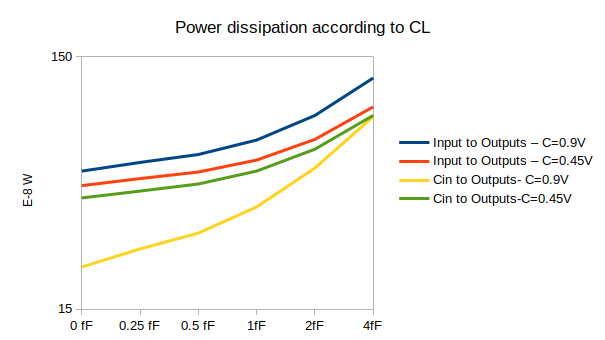}}
\caption{TFA-Power dissipation according to $C_{L}$}
\label{TPCL}
\end{figure}

With a log-log scale (except for $C_{L}$ = 0 fF), Fig. \ref{TI20CL} presents the input to outputs delays according to $C_{L}$. Fig. \ref{TC20CL} presents the same information for $C_{in}$ to outputs delay while Fig. \ref{TPCL} present the evolution of power according to $C_{L}$. Considering the different curves between $C_{L}$ = 0.25fF and $C_{L}$ = 4fF, we may observe that the delay evolution are close to a linear one, with different slopes. Power increases more than linearly according to $C_{L}$.

$C_{in}$ to $C_{out}$ path is through a multiplexer and an inverter while $C_{in}$ to Sum is just through a multiplexer. The inverter restores the signal and has more driving capability than the multiplexer. It explains why the sum delay is more sensitive to capacitive load. Input to $C_{out}$ and Sum paths include the whole circuit. The final inverter delay for $C_{out}$ has a limited impact on the overall delay compared to Sum delay, which explain why these large delays don't increase much when $C_{L}$ is multiplied by 16. Power increases from x2 to x3.

\begin{figure}[htbp]
\centerline{\includegraphics[width=9cm]{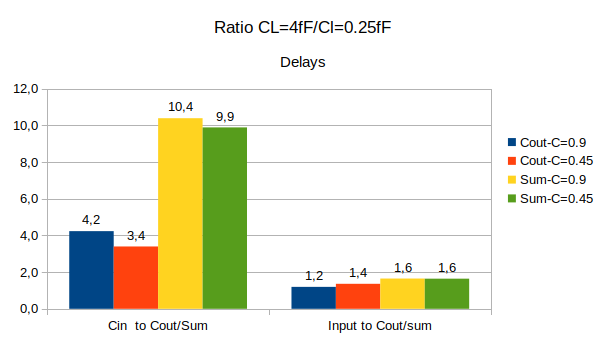}}
\caption{TFA-Delay ratio when $C_{L}$ is multiplied by 16}
\label{TDCLR}
\end{figure}

\begin{figure}[htbp]
\centerline{\includegraphics[width=9cm]{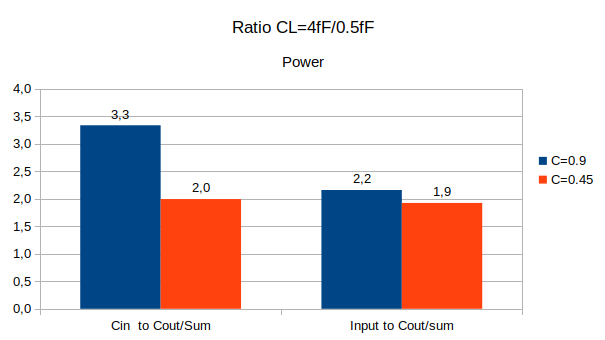}}
\caption{TFA-Power ratio when $C_{L}$ is multiplied by 16}
\label{TPCLR}
\end{figure}

Fig. \ref{TDCLR} presents the ratio delays($C_{L}$ = 4fF)/delays(0.25fF) when $C_{L}$ is multiplied by 16. It is a figure of the slope of the quasi-linear evolution of delays($C_{L}$). For $C_{in}$ to output delays, the sum output is more sensitive to $C_{L}$ than $C_{out}$. It comes from the Sum MUX output that has less driving capability than the $C_{out}$ inverter. Due to the large delays from input to outputs, the impact of $C_{L}$ is limited for these delays. Fig.\ref{TPCLR} presents the power evolution when $C_{L}$ is multiplied by 16. The impact is slightly more important for $C_{in}$ to Outputs than for Input to Outputs as it concerns only MUXes and the final inverter. The $V_{dd}$ inverter consumes more than the $V_{dd}$/2 inverter.

\section{Quaternary Full Adders}
\label{QA}
\begin{figure}[htbp]
\centerline{\includegraphics[width=8cm]{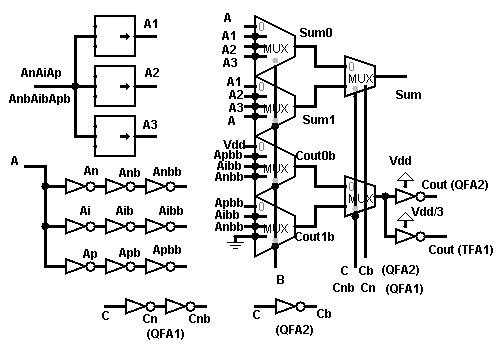}}
\caption{Quaternary Full Adder (MUX approach)}
\label{QFAMUX}
\end{figure}

The common scheme is presented in Fig. \ref{QFAMUX}. The two QFAs only differ by the carry swing. Carry input values are $V_{dd}/3$ (QFA1) and $V_{dd}$ (QFA2). The control of the two MUX2 is shown is Fig. \ref{QFAMUX}. The carry output is obtained by inverters with $V_{dd}/3$ supply (QFA1) or $V_{dd}$ (QFA2)

\subsection{Multiplexer Implementation}
\begin{figure}[htbp]
\centerline{\includegraphics  [width =5 cm]{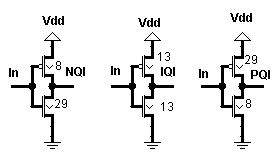}}
\caption{Threshold detectors}
\label{QDEC4}
\end{figure}

\begin{figure}[htbp]
\centerline{\includegraphics[width=8cm]{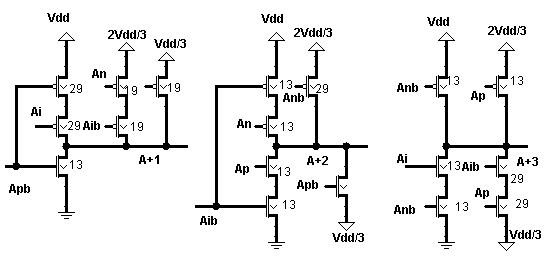}}
\caption{A¹, A² and A³ circuits}
\label{A1A2A3}
\end{figure}

\begin{figure}[htbp]
\centerline{\includegraphics[width=6cm]{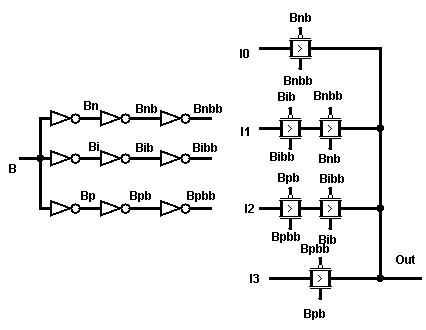}}
\caption{4-input MUX with quaternary control}
\label{QMUXDE}
\end{figure}

The common functional scheme is shown in Fig. \ref{QFAMUX}. The threshold detectors (Fig. \ref{QDEC4}), the circuits A¹, A², A³ (Fig. \ref{A1A2A3}) and the MUX4 (Fig. \ref{QMUXDE}) are similar to those of \cite{R3}. The two final multiplexers are typical binary multiplexers. $\overline{C_{out}}$ is computed from $\overline{C_{out0}}$ and $\overline{C_{out1}}$. A final inverter delivers $C_{out}$. 4-input multiplexers with quaternary control are used (Fig. \ref{QMUXDE}). The three inverters with outputs $B_{nbb}$, $B_{ibb}$ and $B_{pbb}$ operate as buffers because inverters $B_n$ and $B_p$ have poor driving capability. Paper \cite{R3} first uses a quaternary half adder (sum and carry circuits). A second stage computes the final result by adding +1 mod(4) to sum when $C_{in}$=1 and computing $C_{out}$ according to $C_{in}$. We directly computes Sum and $C_{out}$ within a single stage. $C_{in}$ to $C_{out}$ propagation delay is reduced to a MUX2 and final inverter path.

\begin{figure*}[htbp]
\centerline{\includegraphics[width=12cm]{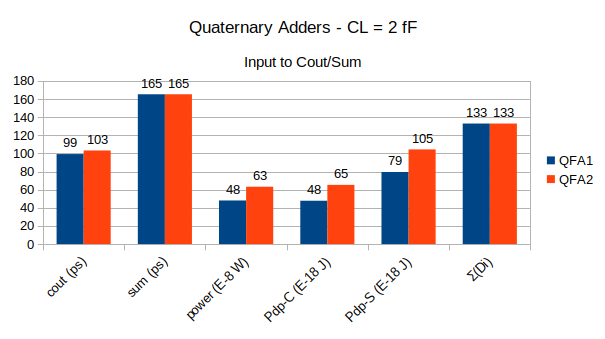}}
\caption{Input to $C_{out}$/Sum performance for QFA1 and QFA2}
\label{QFAI20}
\end{figure*}

\begin{figure*}[htbp]
\centerline{\includegraphics[width=12cm]{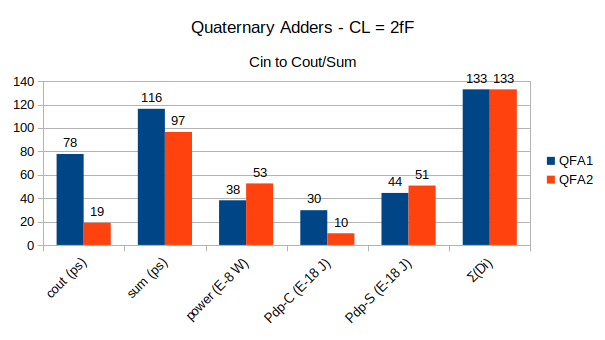}}
\caption{$C_{in}$ to $C_{out}$/Sum performance for QFA1 and QFA2}
\label{QFAC20}
\end{figure*}

\subsection{Performance with a 2 fF capacitive load}
For all simulations, the same input waveforms are used. Extensive simulations have determined that 0$\to$1$\to$2$\to$3$\to$2$\to$1 $\to$0 for input A with $C_{in}$=0 lead to the input to $C_{out}$/Sum worst case delays. Similarly, 0$\to $1 (QFA1)/3 (QFA2) $\to $0 with A=2 and B=1 lead to the $C_{in}$ to $C_{out}$/Sum worst case delays. These configurations are used to evaluate the performance of QFA1 and QFA2. The only difference is the amplitude of the carry swing. 
The performance results are presented in Fig. \ref{QFAI20} and \ref{QFAC20}. These figures provide the data  and allows a direct comparison for for each feature. The significant information is Input to $C_{out}$ (first adder of a CPA), $C_{in}$ to $C_{out}$ (following adders) and $C_{in}$ to Sum (last adder of a CPA).

QFA1 and QFA2 have simular $\Sigma{Di}$. QFA1 has a small advantage in term of power. However, it is outperformed by QFA2 for $C_{in}$ to $C_{out}$ delay, which is the critical delay for a CPA. The situation is the same for PDP. This big advantage comes from the last carry inverter that performs better with a $V_{dd}$ supply than with a $V_{dd}/3$ supply.

\subsection{Delays and power according to capacitive load}
\label{QR}
We now present the performance of QFA1 (0.3V carry swing) and QFA2 (0.9V carry swing). 

\begin{figure}[htbp]
\centerline{\includegraphics[width=9cm]{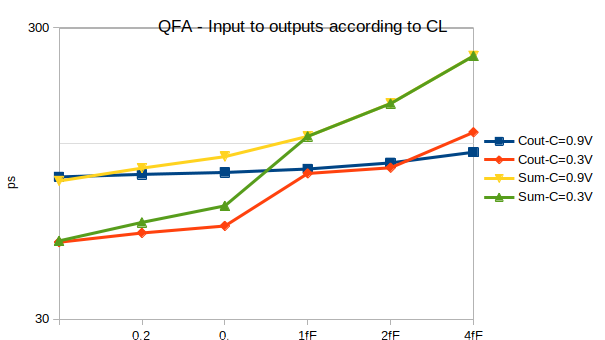}}
\caption{QFA-Input to $C_{out}$/Sum delays according to $C_{L}$}
\label{QI20CL}
\end{figure}

\begin{figure}[htbp]
\centerline{\includegraphics[width=9cm]{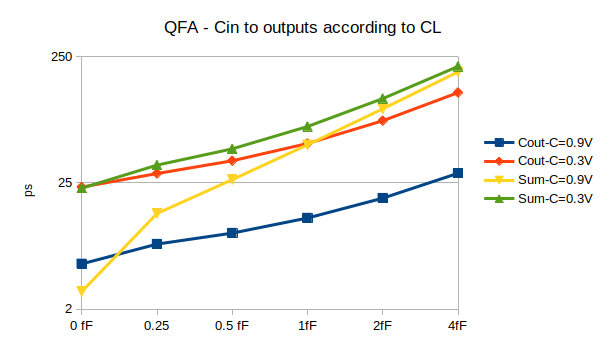}}
\caption{QFA-$C_{in}$ to $C_{out}$/Sum delays according to $C_{L}$}
\label{QC20CL}
\end{figure}

\begin{figure}[htbp]
\centerline{\includegraphics[width=9cm]{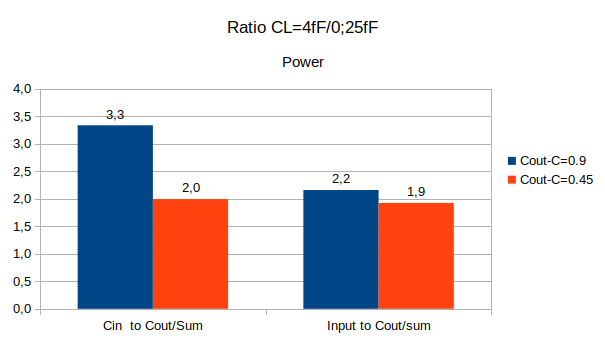}}
\caption{QFA-Power dissipation according to $C_{L}$}
\label{QPCL}
\end{figure}

\begin{figure}[htbp]
\centerline{\includegraphics[width=9cm]{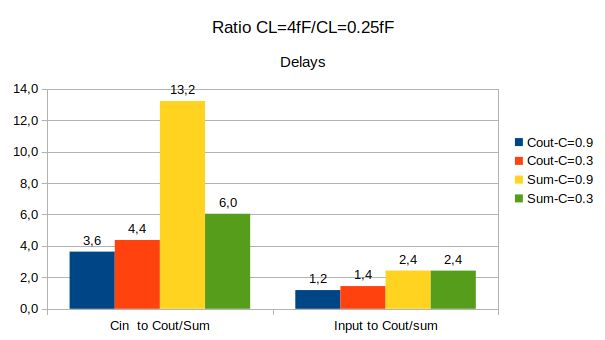}}
\caption{QFA-Delay ratio when $C_{L}$ is multiplied by 16}
\label{QDCLR}
\end{figure}

With a log-log scale, Fig. \ref{QI20CL} presents the input to outputs delays according to $C_{L}$. Fig. \ref{QC20CL} presents the same information for $C_{in}$ to outputs delays while Fig. \ref{QPCL} presents the evolution of power according to $C_{L}$.
Fig. \ref{QDCLR} presents the ratio delays($C_{L}$ = 4fF)/delays(0.25fF) when $C_{L}$ is multiplied by 16. It is a figure of the slope of the quasi-linear evolution of delays($C_{L}$). Fig.\ref{TPCLR} presents the power evolution when $C_{L}$ is multiplied by 16.
The ternary adder and the quaternary adder having the same basic circuit structure, it is not surprising that the conclusions detailed in \ref{TR} are also valid for the quaternary adder. 

\begin{figure}[htbp]
\centerline{\includegraphics[width=9cm]{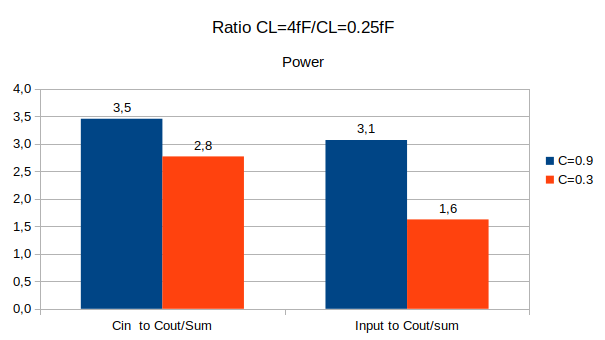}}
\caption{QFA-Power ratio when $C_{L}$ is multiplied by 16}
\label{QPCLR}
\end{figure}

\begin{figure}[htbp]
\centerline{\includegraphics[width=8cm]{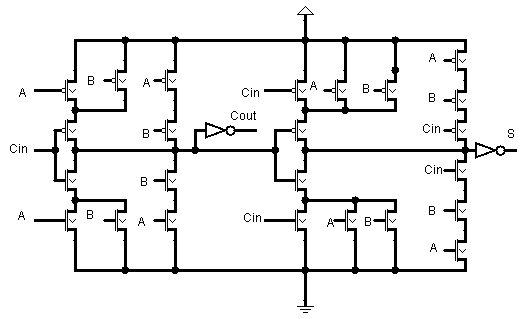}}
\caption{28T Binary Full Adder - BFA2}
\label{28T}
\end{figure}

\begin{figure}[htbp]
\centerline{\includegraphics[width=6cm]{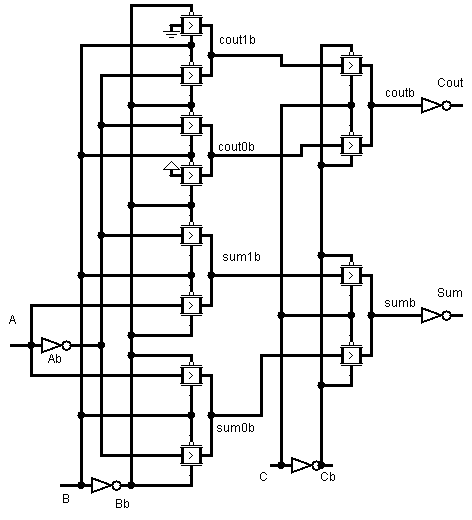}}
\caption{Binary Full Adder-MUX approach (BFA3)}
\label{MUXBFA}
\end{figure}

\section{The Binary Full Adders}
\label{BA}
\subsection{Presentation}
\begin{figure}[htbp]
\centerline{\includegraphics[width=6cm]{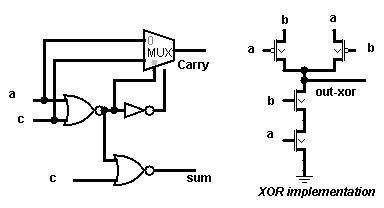}}
\caption{14T Binary Full Adder - BFA1}
\label{14T}
\end{figure}

For the comparison with ternary and quaternary full adders, we consider three different binary adders:
\begin{itemize}
\item The first one is a 14T Full Adder (Fig. \ref{14T}) 
\item The second one is the typical 28T full adder (Fig.\ref{28T})
\item The third one is a MUX-based full adder (Fig. \ref{MUXBFA}) that uses the same circuit style than the ternary and quaternary adders. Using the same circuit style allows a fair comparison.
\end{itemize}
The three binary full adders operate with the same $V_{dd}$ = 0.9V as the quaternary adder. They can also operate with a 0.45V supply, which roughly divide by 4 the dynamic power dissipation. $V_{dd}$ = 0.45V is a too small power supply value to operate with the three levels of a ternary adder or four levels of a quaternary adder.

\begin{figure}[htbp]
\centerline{\includegraphics[width=8cm]{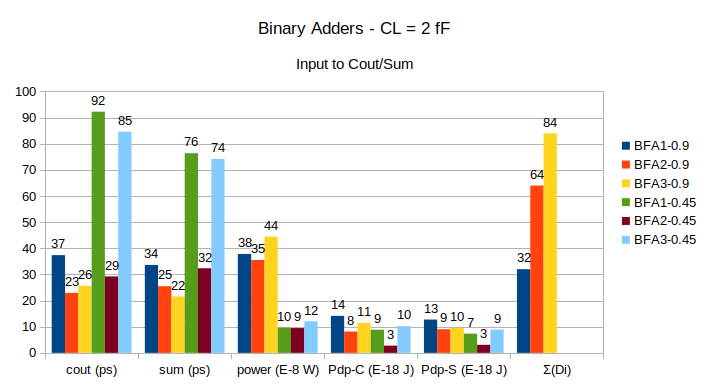}}
\caption{Binary Adders - Input to $C_{out}$/Sum - $C_{L}$ = 2 fF}
\label{BFAI20}
\end{figure}

\begin{figure}[htbp]
\centerline{\includegraphics[width=8cm]{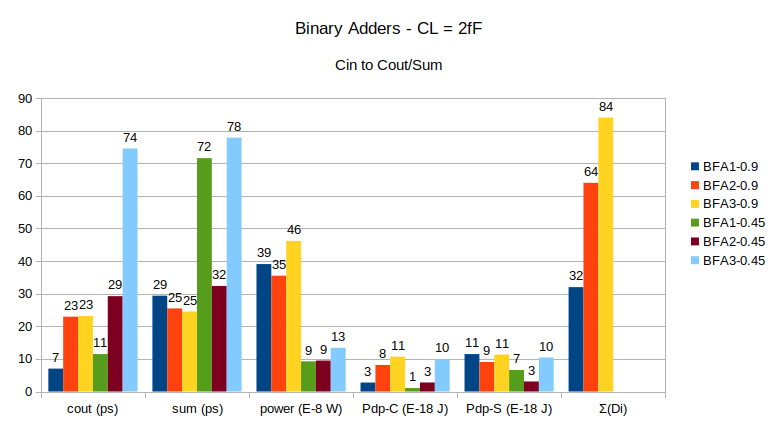}}
\caption{Binary Adders - Input to $C_{out}$/Sum - $C_{L}$ = 2 fF}
\label{BFAC20}
\end{figure}

\subsection{Performance with a 2 fF capacitive load}

Fig. \ref{BFAI20} presents the Input to $C_{out}$/Sum performance with $C_{L}$ = 2 fF. Fig. \ref{BFAC20} presents the $C_{in}$ to $C_{out}$/Sum performance with the same capacitive load. While the MUX-approach (BFA3) is the best approach for ternary and quaternary adders, it is the worst one for binary adder in terms of delays, power and $\Sigma{Di}$. All powers for 0.45 $V_{dd}$ are roughly 1/4 of the powers of 0.9 $V_{dd}$ versions, leading to PDD slightly smaller or equivalent for both $V_{dd}$. For input to $C_{out}$ performance corresponding to the worst case of the first BFA in a CPA, BFA2 is better than BFA1. However $C_{in}$ to $C_{out}$ BFA1 delay, which is the critical delay in a CPA, is about 3x smaller than BFA2 delay both for 0.9V and 0.45V $V_{dd}$. The $C_{in}$ to Sum delays, which is critical for the last stage of a CPA, are close for BFA1 and BFA2 ($V_{dd}$ = 9V) and x2.25 greater for BFA1 ($V_{dd}$ = 0.45V), but this is conterbalanced by the x3 smaller $C_{in}$ to $C_{out}$ delay for TFA1. 

BFA1 is globally the most efficient binary adder in terms of delays, PDP and $\Sigma{Di}$ for the two different power supplies.

\subsection{Delays and power according to capacitive load}
\label{BR}
\begin{figure}[htbp]
\centerline{\includegraphics[width=9cm]{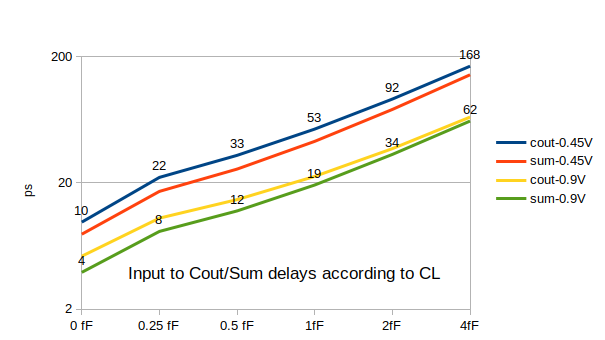}}
\caption{BFA1-Input to $C_{out}$/Sum delays according to $C_{L}$}
\label{BI20CL}
\end{figure}

\begin{figure}[htbp]
\centerline{\includegraphics[width=9cm]{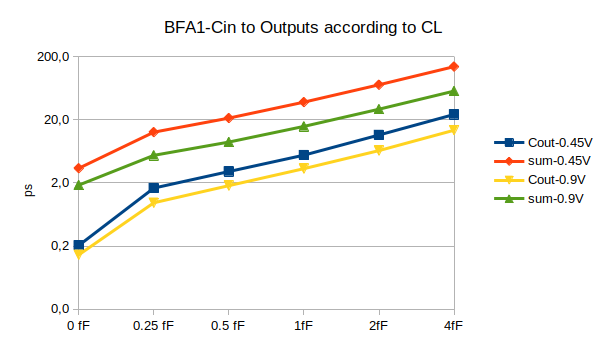}}
\caption{BFA1-$C_{in}$ to $C_{out}$/Sum delays according to $C_{L}$}
\label{BC20CL}
\end{figure}

\begin{figure}[htbp]
\centerline{\includegraphics[width=9cm]{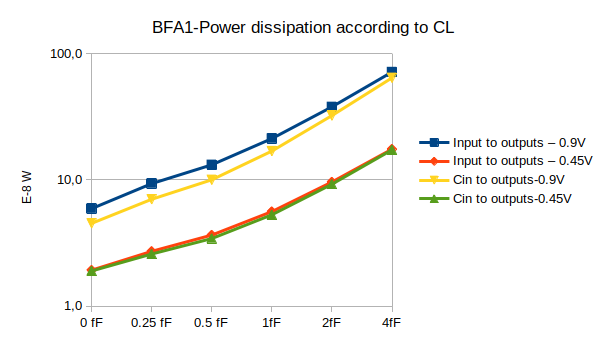}}
\caption{BFA1-Power dissipation according to $C_{L}$}
\label{BPCL}
\end{figure}

\begin{figure}[htbp]
\centerline{\includegraphics[width=9cm]{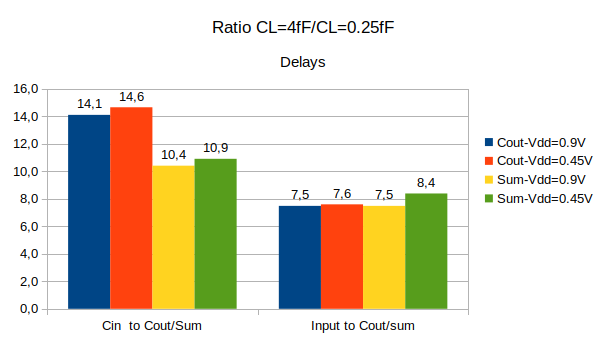}}
\caption{BFA-Delay ratio when $C_{L}$ is multiplied by 16}
\label{BDCLR}
\end{figure}

\begin{figure}[htbp]
\centerline{\includegraphics[width=9cm]{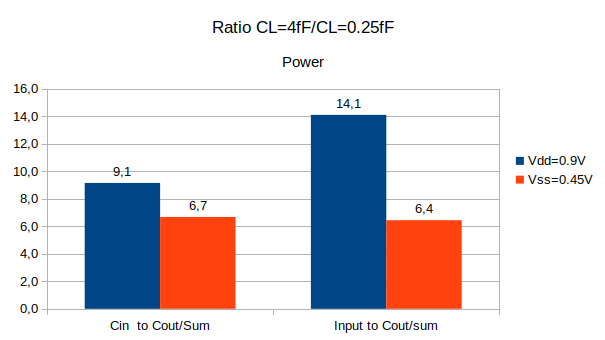}}
\caption{BFA-Power ratio when $C_{L}$ is multiplied by 16}
\label{BPCLR}
\end{figure}

We now present the performance of BFA1 according to capacitive loads and temperature.
With a log-log scale, Fig. \ref{BI20CL} presents the input to outputs delays according to $C_{L}$. Fig. \ref{BC20CL} presents the same information for $C_{in}$ to outputs delays while Fig. \ref{BPCL} presents the evolution of power according to $C_{L}$.
Fig. \ref{BDCLR} presents the ratio delays($C_{L}$ = 4fF)/delays(0.25fF) when $C_{L}$ is multiplied by 16. It is a figure of the slope of the quasi-linear evolution of delays($C_{L}$). Fig.\ref{BPCLR} presents the power evolution when $C_{L}$ is multiplied by 16.
We still have a quasi linear evolution of delay and power according to $C_{L}$. However, the binary adder structure is different of the m-valued adder structures: there is one MUX for $C_{out}$, but not a series of MUXes as in the Sum output of ternary and quaternary adders. Globally, the binary adder is more sensitive to capacitive loads than the ternary and quaternary ones.

\section{Comparing 6-bit,4-trit and 3-qdigit CPAs}
Results provided in \ref{TR}, \ref{QR} and \ref{BR} allow a detailed comparison of the performance of the different adders to be used in a Carry-Propagate Adder. 
The most significant information is to compare CPAs computing the same amount of information. It is strictly the case for 6-bit and 3 quit CPAs. 4-trit input corresponds to 6.34 bits, which corresponds about to 6\% more information than 6-bit or 3 quit. 

Several 4-trit CPAs have been presented in the literature \cite{R3}, \cite{R10}, \cite{R11} and \cite{Jaber20213}.

\begin{figure*}[htbp]
\centerline{\includegraphics[width=16cm]{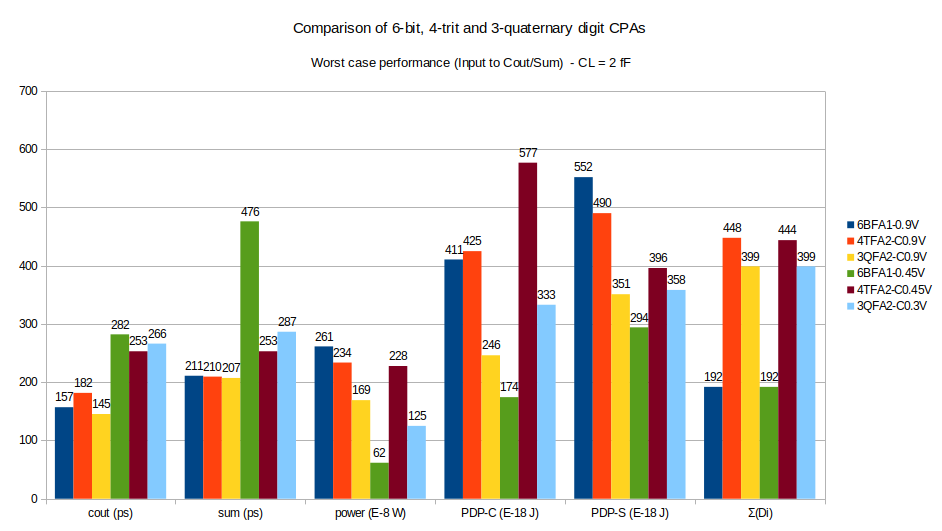}}
\caption{Comparing 6-bit, 4-trit and 3-qdigit CPAs with $C_{L}$ = 2 fF}
\label{C643}
\end{figure*}

Fig. \ref{C643} compares the performance of these three CPAs with two variants: the ternary one uses 0-$V_{dd}$/2 and 0-$V_{dd}$ carry swing, the quaternary one uses 0-$V_{dd}$/3 and 0-$V_{dd}$ carry swing and the binary one uses $V_{dd}$ and $V_{dd}$/2 power supplies. The simulation have been done with a $C_{L}$ = 2 fF capacitive load and T = 25°C temperature. Other loads or temperatures would not change the results of the comparisons.
From Fig. \ref{C643}, the following conclusions can be deduced:
\begin{itemize}
\item While the binary CPA uses more full adders, its estimated chip area is half the chip area of the ternary and quaternary CPAs.
\item The ternary and quaternary CPAs have less propagation delays when using full carry swing than when using $V_{dd}$/2 or $V_{dd}$/3 carry swing
\item The 0.45 $V_{dd}$ binary CPAs has the smallest power dissipation, from 1/2 to 1/4 power dissipation of the other CPAs. While its input to sum delay is the worst one, this CPA has the lowest PDP both for sum and carry outputs.
\item The quaternary CPA has a small advantage for delays with full carry swing, but the values are closed.
\end{itemize}
While ternary and quaternary CPAs have less full adders, they suffer from the large chip area and don't provide significant advantages in term of delays. The best CPA is the binary one with $V_{dd}$ = 0.45V supply. Reducing power supply is possible with binary circuits, but is not possible with ternary and quaternary circuits as they need a larger $V_{dd}$ to handle the different voltage levels.

\section{Concluding remarks}
We have detailed the performance of binary, ternary and quaternary full adders that are probably close to the most performant ones. We have shown that two options are possible for these adders.
\begin{itemize}
\item For ternary and quaternary adders, we used two carry swings. The first one correspond to 0-1 logical values, i.e. 0-$V_{dd}$/2 for the ternary adder and 0-$V_{dd}$/3 for the quaternary adder. The second one uses the 0-$V_{dd}$ carry swing for both adders, as carry values are always 0-1 logical values for any radix used for addition. It turns out that full carry swing reduces significantly carry propagation delays with a small power increase.
\item For binary adders, we use both 0.9V and 0.45V power supplies. The smallest $V_{dd}$ value reduces significantly power (/4 factor), which leads to reduced PDP with a small increase in delays
\end{itemize}
The different adders are used in CPAs computing the same amount of information.
In CPAs, carry propagation is the critical delay. The critical delay paths are similar for the ternary and quaternary adders. For the binary adder, it consists in a NOR gate and a MUX. With 6, 4 and 3 adders in the binary, ternary and quaternary cases, the ternary and quaternary adders should benefit from the reduced number of carry paths. It turns out that this is not the case as input to  carry delays are close (they are not in the ratio 6/4/3). Input to Sum delays are also close with 0.9V $V_{dd}$ (binary) and 0.9V swing (ternary and quaternary).

The only figure for which 3-quit CPAs shows a small advantage is input to output delays with 0.9V carry swing. The ternary and quaternary CPAs are outperformed by the binary CPA with 0.45V supply in terms of power and PDP.

CPAs are circuits for which moving from binary to ternary or quaternary N-digit CPAs is simple: just replace the binary full adders by ternary or quaternary full adders. Moving from a N*N digit binary multiplier to a N*N digit ternary or quaternary multiplier is not so simple. Combinational multipliers using Wallace tree reduction circuits (or equivalent ones) uses both 1-digit multipliers and adders. Ternary and quaternary multiplications generates both product and carry values while binary multiplication (AND gate) only generates 1 bit product.

\end{document}